# Tuning bad metal and non-Fermi liquid behavior in a Mott material: rare earth nickelate thin films

Evgeny Mikheev[1], Adam J. Hauser[1], Burak Himmetoglu[1], Nelson E. Moreno[1], Anderson Janotti[1], Chris G. Van de Walle[1], and Susanne Stemmer[1*]

[1]Materials Department, University of California, Santa Barbara, CA 93106-5050, U.S.A.

**Abstract**

Resistances that exceed the Mott-Ioffe-Regel limit, known as bad metal behavior, and non-Fermi liquid behavior are ubiquitous features of the normal state of many strongly correlated materials. Here we establish the conditions that lead to bad metal and non-Fermi liquid phases in $NdNiO_3$, which exhibits a prototype, bandwidth-controlled metal-insulator transition. We show that resistance saturation is determined by the magnitude of the Ni $e_g$ orbital splitting, which can be tuned by strain in epitaxial films, causing the appearance of bad metal behavior under certain conditions. The results shed light on the nature of a crossover to non-Fermi liquid metal phase and provide a predictive criterion for strong localization. They elucidate a seemingly complex phase behavior as a function of film strain and confinement and provide guidelines for orbital engineering and novel devices.

# Introduction

Mott metal-insulator transitions (MITs) are key to some of the most fascinating topics in materials physics, such as the pathways from a doped Mott insulator to a high-temperature superconductor *(1)*, and the feasibility of electronic devices that utilize switchable MITs. The rare-earth nickelates ($R$NiO$_3$, where $R$ = trivalent rare earth ion) exhibit a prototype bandwidth-controlled MIT *(2, 3)*. $R$NiO$_3$ films have recently attracted renewed interest due to predictions that orbital engineering can promote a Fermi surface that resembles that of the cuprate high-temperature superconductors *(4, 5)*. Furthermore, recent discoveries in $R$NiO$_3$ films point to strikingly similar physics as found in unconventional superconductors. Non-Fermi liquid behavior *(6)* and pseudogap phases *(7)* indicate a continuous, bandwidth-driven quantum phase transition between a paramagnetic metal and an antiferromagnetic insulator. The $R$NiO$_3$ family has also been discussed as a new class of "bad metals" *(8)*, in the sense that their resistivity escalates above the semiclassical Mott-Ioffe-Regel limit and does not saturate at high temperatures *(9)*. Bad metal behavior and non-Fermi liquids are essential yet poorly understood features of the phase diagrams of unconventional superconductors *(10)*. However, neither "bad metals" nor their counterparts, metals that exhibit resistance saturation, are understood – in both cases materials enter a regime where classical Boltzmann theory should no longer apply *(11-14)*.

In this work, we show that the rare earth nickelates are not "bad metals": accounting for resistivity saturation is key to correctly describe their electrical transport behavior. The resistivity saturation limit is, however, highly sensitive to the degree of $e_g$ orbital polarization, leading to resistances that exceed the semiclassical Mott-Ioffe-Regel limit. Furthermore, accounting for saturation clarifies many aspects of the epitaxial strain-film thickness phase behavior and the quantum critical point in the $R$NiO$_3$ system. In particular, an abrupt crossover between classic Landau Fermi liquid (LFL) and non-Fermi liquid (NFL) metallic regimes occurs with the suppression of the temperature-driven MIT. The metallic phase is a LFL in all cases where a robust MIT is present. We also clarify the conditions leading to strong localization in this system, namely a second, disorder-driven MIT: it appears when the 0-K resistivity approaches the saturation resistance. Phase diagrams are developed that can serve as practical guidelines for stabilizing robust MITs in ultrathin nickelate films and that identify new opportunities for control of MITs in general.

**Results**

NdNiO$_3$ thin films with thicknesses ranging between 4 and 15 unit cells (u.c.'s) were grown on substrates chosen to obtain a wide range of epitaxial strains (given in parentheses): YAlO$_3$ (-3.58 %), LaAlO$_3$ (-1.20 %), NdGaO$_3$ (+0.86 %), (LaAlO$_3$)$_{0.3}$(Sr$_2$AlTaO$_6$)$_{0.7}$ (LSAT, +0.93%), SrTiO$_3$ (+1.72%) and DyScO$_3$ (+2.96%). Figure 1 shows their electrical resistivities as a function of temperature. Several different types of behavior can be discerned, including temperature-driven MITs, as well as films that are metallic or insulating at all temperatures, respectively.

We first focus on describing the metallic states. $R$NiO$_3$s are often understood to be NFLs *(6, 15-19)*, for which the temperature ($T$) dependence of the resistivity follows a power law with an exponent $n < 2$:

$$\rho_{NFL}(T) = \rho_0 + AT^n, \qquad (1)$$

where $\rho_0$ is the residual resistivity and $A$ is a measure of the strength of electron-electron scattering. For a LFL, $n = 2$. However, Eq. (1) often describes the experimental data only in a limited temperature range *(20)*. As will be shown here, the temperature dependence of the resistivity in the metallic phase is completely described when we account for resistivity saturation ($\rho_{SAT}$):

$$\rho^{-1}(T) = \rho_{NFL}^{-1}(T) + \rho_{SAT}^{-1} \qquad (2)$$

Eq. (2), in which $\rho_{SAT}$ acts as a parallel resistor, is known to apply to a wide range of materials, but the origins of $\rho_{SAT}$ remain a subject of significant debate *(10-12, 21-26)*. $\rho_{SAT}$ is often linked to high resistances that approach the Mott-Ioffe-Regel ($\rho_{MIR}$) limit *(23)*, which is the semi-classical upper bound for coherent transport in a metal, when the carrier mean free path $l$ approaches the interatomic spacing $a$:

$$\rho_{MIR} = \frac{3\pi^2 \hbar}{q^2 k_F^2 a}, \qquad (3)$$

where $\hbar$ is the reduced Planck's constant, $q$ the elementary charge, and $k_F$ is the Fermi wavevector. For some materials, $\rho_{MIR}$ estimated from Eq. (3) is similar to the observed $\rho_{SAT}$ *(10)*. Others, including some of the unconventional superconductors *(22)*, are characterized by saturation at values much higher than $\rho_{MIR}$ calculated by Eq. (3), or even a non-saturating $\rho$, and these materials have become known as "bad metals" *(9)*. Quantum Monte Carlo and dynamical mean field theory calculations indicate that resistances can easily exceed $\rho_{MIR}$ calculated from Eq. (3) for certain materials *(13, 14)*.

For NdNiO$_3$ films, Eq. (2) is remarkably successful in describing the metallic state. This can be seen from the dashed lines in Fig. 1, which are fits to Eq. (2), and it is also particularly obvious in d$\rho$/d$T$, as shown for one example in Fig. 2(a). A key feature of Eq. (2) is the downturn in d$\rho$/d$T$ at high $T$, which is due to the increasing contribution from $\rho_{SAT}$. Eq. (2) describes the entire metallic state with a single *T*-independent exponent *n* (see the inset in Fig. 2(a)). Additional examples, including for LaNiO$_3$ films, are shown in the Supplementary Material.

While the exponent *n* was an adjustable parameter in the fits, only two different values were obtained across the entire sample set (Fig. 2(b)): for all films displaying a robust temperature-driven MIT *n* = 2, indicating a classic LFL. For films that are metallic at all temperatures (MIT completely suppressed) and some films with a weak MIT at very low *T*, $n \approx 5/3$, indicating a NFL regime. For example, on YAlO$_3$, all but the 4 u.c. film are metallic at all temperatures, and NFLs. The 4 u.c. film shows an MIT and recovers LFL behavior. We note that *not* including $\rho_{SAT}$ *(6, 15-18)*, would have resulted in interpreting the metallic state as a NFL even in case of a LFL film, with apparent changes in the exponent *n* for different temperature ranges, as illustrated in Fig. 2(a). For example, *n* = 1 is assumed and $\rho_{SAT}$ is neglected for the dashed orange line in Fig. 2(a). For the purpose of the final fits shown in Fig. 1 the exponent *n* was fixed at 2 or 5/3 (by closest value). The changes in the results were minimal, but a fixed exponent allows for more reliable comparisons across the series for the slope *A* (see Supplementary Information).

Figure 2(c) shows $\rho_{SAT}$ and $\rho(0)$, the metallic resistivity extrapolated to *T* = 0 K, as extracted from the fits, as a function of film thickness. Following Eq. (2):

$$\rho^{-1}(0) = \rho_{NFL}^{-1}(0) + \rho_{SAT}^{-1} = \rho_0^{-1} + \rho_{SAT}^{-1} \,. \qquad (4)$$

We note that $\rho_0 \ll \rho_{SAT}$ for most films, so $\rho(0) \sim \rho_0$. At low thicknesses, $\rho_0$ sharply increases, which is the ubiquitously observed rise of the resistivity in ultrathin nickelates *(27-30)*. In contrast, $\rho_{SAT}$ is essentially thickness-independent. $\rho_{SAT}$ does depend, however, on the magnitude of the epitaxial strain as determined by the substrate. This is further illustrated in Fig. 2(d), where $\rho_{SAT}$, $\rho_0$ and $\rho(0)$ are shown as a function of in-plane strain $\varepsilon_{xx}$ for the 15 u.c films. $\rho_{SAT}$ increases for both compressive and tensile strains. A larger increase is observed on the tensile side, where an enhancement by a factor of ~ 4 is found for the case of a metallic film on SrTiO$_3$. Also shown in Fig. 2(d) as a dashed line is an estimate of $\rho_{MIR}$ using Eq. (3) with $k_F^3 = 3\pi^2 N$ and a carrier density $N \approx 10^{22}$ cm$^{-3}$, which gives $\rho_{MIR} \approx 0.5$ mΩcm *(8)*. Only for small strains is $\rho_{SAT}$ similar to $\rho_{MIR}$ calculated by this estimate. Analysis of LaNiO$_3$ films *(27)* reveals a $\rho_{SAT}$ as high as 4 mΩcm (see Supplementary Information).

The conditions yielding LFL, NFL, and MITs are summarized in Figs. 3 and 4. Figure 3(b) shows the phase behavior as a function of strain and film thickness (*t*), and we distinguish four types of $\rho$-*T* curves (see examples Fig. 3(a)): paramagnetic NFL metal at all temperatures (PM(NFL), light blue color), a FL metal at high temperatures with a sharp, hysteretic transition to an antiferromagnetic Mott insulator near 150 K (PM(FL)↔AFI, yellow), a NFL metal with a strongly suppressed MIT (PM(NFL)↔AFI, dark blue), and insulating at all measured *T* (AFI, red). An important feature in Fig. 3 is the pronounced curvature of the transition boundaries. This is a result of similar transport types shifting to lower thickness with increasing compressive strain. For example, the PM(FL)↔AFI region (sharp hysteretic transition, *n* = 2) occurs at *t* > 6 u.c. for tensile strain, at 6 u.c only for films LaAlO$_3$ and at 4 u.c. only for films on YAlO$_3$ (largest compressive strain). Interestingly, similar trends are seen for the occurrence of PM(NFL), PM(NFL)↔I and the "Anderson insulator" (insulating behavior at all temperatures, red region).

Figure 4 shows the strain-temperature phase diagram. The transition boundaries shown are consistent with Fig. 3. The curvature of the boundaries in Fig. 3 is a result of a lateral shift (parallel to the strain axis) of the entire phase diagram with NdNiO$_3$ film thickness.

**Discussion**

**Tunable Bad Metal**

A key result is that NdNiO$_3$ clearly exhibits resistance saturation in the high temperature limit. It is thus not a "bad metal" only in the sense that the saturation resistance exceeds the resistance predicted by Eq. (3) but not in the sense that its resistance escalates without saturation (as in the curprates). In the following we discuss the origins of this behavior and its correlation to the electronic structure of this system.

Several theoretical studies have pointed to the importance of orbital degeneracy and of specific scattering mechanisms in determining $\rho_{SAT}$ *(25, 31)*. Transport calculations that include interband scattering show that this produces a new conducting channel, whose magnitude is proportional within first order to the interband spacing $\Delta$ *(24, 25)*. In this theory, the interband currents act as a parallel conducting channel that reduces the resistance, leading to a saturating resistance as described by Eq. (2), with $\rho_{SAT} \sim |\Delta|$. The behavior of $\rho_{SAT}$ as a function of strain in NdNiO$_3$ thin films can then be rationalized as a consequence of the e$_g$ band splitting. In the rare earth nickelates, two Ni e$_g$ bands that cross the Fermi level. These are derived from orbitals having $x^2-y^2$ and $3z^2-r^2$ symmetry, which are degenerate in the unstrained, bulk material. Experiments have shown that epitaxial strain lifts the degeneracy and cause orbital polarization, with tensile strains lowering the energy $E$ of the $x^2-y^2$ orbitals and compressive strains that of the $3z^2-r^2$ orbitals *(32-34)*. Figure 5 shows the magnitude of the orbital splitting $|\Delta| = E(3z^2-r^2) - E(x^2-y^2)$ for NdNiO$_3$ as a function of strain, as estimated by density functional theory. With increasing epitaxial strain the orbital polarization increases, in keeping with these prior experimental and theoretical findings *(32-34)*, as does $\rho_{SAT}$. Moreover, we find that the electron-electron scattering strength [$A$ in Eq. (1)] follows a similar trend with strain as $\rho_{SAT}$ (see Supplementary Information). This trend may also be an indication of increasing importance of interband scattering, which can lead to resistance from electron-electron scattering, in addition to Umklapp processes *(35)*. The strain-tunable $\rho_{SAT}$ allows us to establish an understanding of how electronic structure determines $\rho_{SAT}$ and thereby (the degree of) bad metal behavior. Specifically, $\rho_{SAT} \propto |\Delta| = E(3z^2-1)-E(x^2-y^2)$, the orbital polarization (see inset in Fig. 5(b)). Thus, a large orbital splitting causes $\rho_{SAT}$ to rise above the ~ 0.5 mΩcm value predicted by Eq. (3) and the appearance of bad metal behavior, in the sense that $\rho$ exceeds $\rho_{MIR}$.

Calculations for the cuprates predict that these compounds also saturate (as appears to be confirmed in the experiment *(10)*), but that $\rho_{SAT}$ is very large due to the fact that only a single $x^2-y^2$ orbital band crosses the Fermi level *(31, 36)*, in contrast to the nickelates studied here, and due to strong electron correlations. A large $\rho_{SAT}$ makes the second term in Eq. (2) small and causes a non-saturating resistance. The experimental results here confirm the importance of the orbital degeneracy: as we lift the degeneracy towards a more cuprate-like Fermi surface with a single band, $\rho_{SAT}$ increases.

The results also show that while $\rho_{SAT}$ is sensitive to the degree of orbital polarization, it is relatively insensitive to disorder. This can be seen from the very different behaviors of $\rho_{SAT}$ and $\rho_0$ with decreasing film thickness - $\rho_0$ sharply increases, presumably due to the increased scattering by the surface, while $\rho_{SAT}$ stays approximately constant (see Fig. 2(c)).

Using the data shown in Fig. 2(c) predictions can be made when strong localization occurs in this system. In particular, films become insulating at all temperatures when $\rho(0) \approx \rho_{SAT}$, in other words, $\rho_0$ becomes so large that essentially at all temperatures the first term in Eq. (2) is small compared to the second term, and the resistance is dominated by $\rho_{SAT}$. Such a material clearly cannot be a metal anymore, and films are insulating at all temperatures. The black diamonds in the phase diagram shown in Fig. 3(b) are predictions for the critical thickness for this transition, obtained by extrapolating $\rho(0)$ to the point where it intersects with $\rho_{SAT}$. They agree with the experimental transition within one u.c. Since $\rho_0$ contains the effects from disorder, $R$NiO$_3$ films that are insulating at all temperatures are strongly localized due to disorder, as has also been suggested in the literature *(27, 28, 37)*. We loosely term this an "Anderson insulator", although correlations presumably play a role and the insulator may be magnetic *(38)*. The criterion ($\rho(0) \approx \rho_{SAT}$) established here has an interesting implication, namely that the Anderson insulating state is *tunable* in a similar fashion as $\rho_{SAT}$. Specifically, materials with a small $\rho_{SAT}$ will require larger $\rho_0$'s to become insulating (recall that $\rho_{SAT}$ acts to decrease the overall resistance) and can be considered more disorder-tolerant, all other things being equal. Unlike $A$ and $\rho_{SAT}$, however, $\rho_0$ does depend on the sign of the strain and not only on its magnitude. This is due to $\rho_0$ being a function of the size of the Fermi surface(s) *(39)*, which change(s) with strain *(33)*. Tensile strained films, with their larger $\rho_0$, fulfill $\rho(0) \approx \rho_{SAT}$ at larger thicknesses than compressively strained films (see data on SrTiO$_3$ in Fig. 2(c)).

**Non-Fermi Liquid Behavior**

The phase diagrams shown in Figs. 3 and 4 illustrate the two different types of MIT: one is a 0-K quantum phase transition between NFL and an antiferromagnetic insulator (blue-yellow region crossover), the other is the disorder driven transition to the Anderson insulator (yellow-red region crossover), when $\rho(0) \approx \rho_{SAT}$. The quantum phase transition as a function of strain has been documented in relatively thick $R$NiO$_3$ films *(6, 40-42)*. This nature of this MIT has been discussed in the literature, as being driven by Fermi surface nesting and a spin density wave, which promotes the insulating state *(33, 38, 43, 44)*, or in terms of bandwidth and charge-transfer energy *(6, 40, 45)*. The insulating phase at low thicknesses has also been linked to the stabilization of the spin density wave order, acting similarly to tensile strain *(32, 33)*. This work establishes that the location of the quantum phase transition is highly sensitive to both strain and confinement. This is reflected in the curved phase boundaries in Fig. 3 and, equivalently, in the continuous shift of the entire phase diagram towards compressive strain at low thickness in Fig. 4. Moreover, at high tensile strains, the Mott MIT and the disorder driven Anderson transition are brought in increasingly close proximity to each other. The dual nature of the driving forces promoting the insulating transition should be an important consideration in interpreting experiments, such as the magnetism *(28)*. Using strain, the two transitions can be decoupled, e.g. as in $NdNiO_3$ grown on $LaAlO_3$.

The results further have intriguing implications for the nature of the NFL-LFL crossover in the metallic phase, which is abrupt and coincides with the suppression of the temperature-driven MIT. It is noteworthy that the $n$ robust against disorder, which increases (according to $\rho_0$) with decreasing film thickness (see Figs. 2(b) and (c)). This is contrary to expectations of NFL driven by spin fluctuations, which should yield $n$ highly sensitive to disorder *(46)*. It is furthermore constant across all NFL phases observed here. Interestingly, $n \approx 5/3$ has also been observed for $PrNiO_3$ and $EuNiO_3$ under pressure when the temperature-driven MIT is suppressed *(47, 48)*, as well as in overdoped cuprates *(49, 50)*. This points to a common origin that requires further theoretical investigations. Specifically, future studies should address the question if a quantum critical point or a distinct NFL phase are the origin of the NFL metal. The results emphasize the need for a quantitative treatment of electronic structure and also, in keeping with earlier suggestion *(4, 5)*, that the nickelates are a fertile ground to investigate aspects of the normal states of the cuprates.

**Applications**

The phase diagram in Fig. 3 can be used as a guide for the design of future devices based on controlling the MIT of the nickelates. The M(NFL)↔I pocket at compressive strains at low film thicknesses could be useful for electrostatic control *(51-55)*, which requires thin films to make carrier density modulation feasible, while still permitting a sharp MIT with many orders of magnitude of resistance change. Strain-control of the MIT may also be of interest for low-voltage digital switches *(56)*. On the tensile side of the phase diagram, the sharp M(NFL)↔I/I boundary is noteworthy as a small amount of strain, controlled by a piezoelectric, can have a large effect on resistivity, at the temperatures relevant for practical devices. In particular, Fig. 3 shows a strain-tunable transition between the Anderson insulator ($\rho(0) = \rho_{SAT}$) and the Mott-MIT.

## Materials and Methods

Films were grown by RF magnetron sputtering in a 95% Ar/5% $O_2$ mixture, a total pressure of 9 mTorr and a sputter power of 15 W. Details of how the growth conditions were optimized, the films' structure and chemical composition, and the quantification of the mismatch strains have been reported elsewhere *(42)*. Methods used to characterize the films included high-resolution x-ray diffraction, scanning transmission electron microscopy, and Rutherford Backscattering Spectrometry, as described in ref. *(42)*. Films of a given thickness were deposited simultaneously on the different substrates. The magnitude of the in-plane strain was calculated as $\varepsilon_{xx}=(a_\parallel-a_0)/a_0$, where a∥ is the measured in-plane lattice constant and $a_0$ is the unstrained (intrinsic) lattice parameter, which was extracted from the x-ray diffraction as described in ref. *(42)*. The resistivity was determined from mesurements in a Van der Pauw configuration with Ni(20 nm)/Au(300 nm) Ohmic contacts, between 2 and 300 K using a Quantum Design Physical Properties Measurement System (PPMS).

Electronic structure calculations are performed using the projector-augmented wave (PAW) formalism *(57)* as implemented in the Quantum ESPRESSO package *(58)*. The electronic wavefunctions are expanded up to a kinetic energy cut-off of 50 Ry. The Brillouin Zone integrations are performed on a 8×8×8 special k-point grid, and a Methfessel-Paxton smearing *(59)* of the Fermi-Dirac distribution function with a smearing width of 0.01 Ry. We use the generalized gradient approximation for the exchange-correlation functional (PBE) *(60)*. The crystal structure is constrained to a tetragonal unit cell with the *ab* plane fixed to the lattice constant of the substrate, while the *c* lattice parameter is allowed to relax. For the calculation of crystal field splittings, we construct maximally localized Wannier functions *(61)* and a real space Hamiltonian in the Wannier function basis. The splitting between the diagonal elements of the real-space Hamiltonian of the $e_g$-like Wannier functions for different strain configurations is interpreted as crystal-field splitting for the $e_g$ states, defined as $\Delta = H(z^2, z^2) - H(x^2-y^2, x^2-y^2)$, with H representing the real space Hamiltonian in Wannier function basis. While the PBE functional and the tetragonal unit cell do not reproduce the insulating and E'-type anti-ferromagnetic ordering of bulk $NdNiO_3$, they provide the correct qualitative behavior and trends for the dependence of $\Delta$ as a function of strain. In addition, we expect the metallic solution to provide a better description of the electronic properties of metallic $NdNiO_3$ films, compared to that of bulk $NdNiO_3$ within PBE.

**Acknowledgments:** The authors gratefully acknowledge Jim Allen for helpful discussions and critically reading the manuscript, and Junwoo Son for the $LaNiO_3$ data. The authors also thank Leon Balents for discussions. **Funding:** This work was supported in part by FAME, one of six centers of STARnet, a Semiconductor Research Corporation program sponsored by MARCO and DARPA and by the U.S. Army Research Office (grant nos. W911-NF-09-1-0398, W911NF-14-1-0379, and W911-NF-11-1-0232). B.H. was supported by the Office of Naval Research, grant No. N00014-12-1-0976. A. J. H. acknowledges support through an Elings Prize Fellowship of the California Nanosystems Institute at University of California, Santa Barbara. The work made use of central facilities of the UCSB MRL, which is supported by the MRSEC Program of the National Science Foundation under Award No. DMR-1121053. The work also made use of the UCSB Nanofabrication Facility, a part of the NSF-funded NNIN network. **Author Contributions:** E.M., N.E.M., and A.J.H. performed the film growth and electrical measurements. E.M. analyzed the data. B. H. and A. J. carried out the electronic structure calculations. E.M. and S.S. wrote the manuscript and all authors commented on it. **Competing financial interests:** The authors declare no competing financial interests

**Figures**

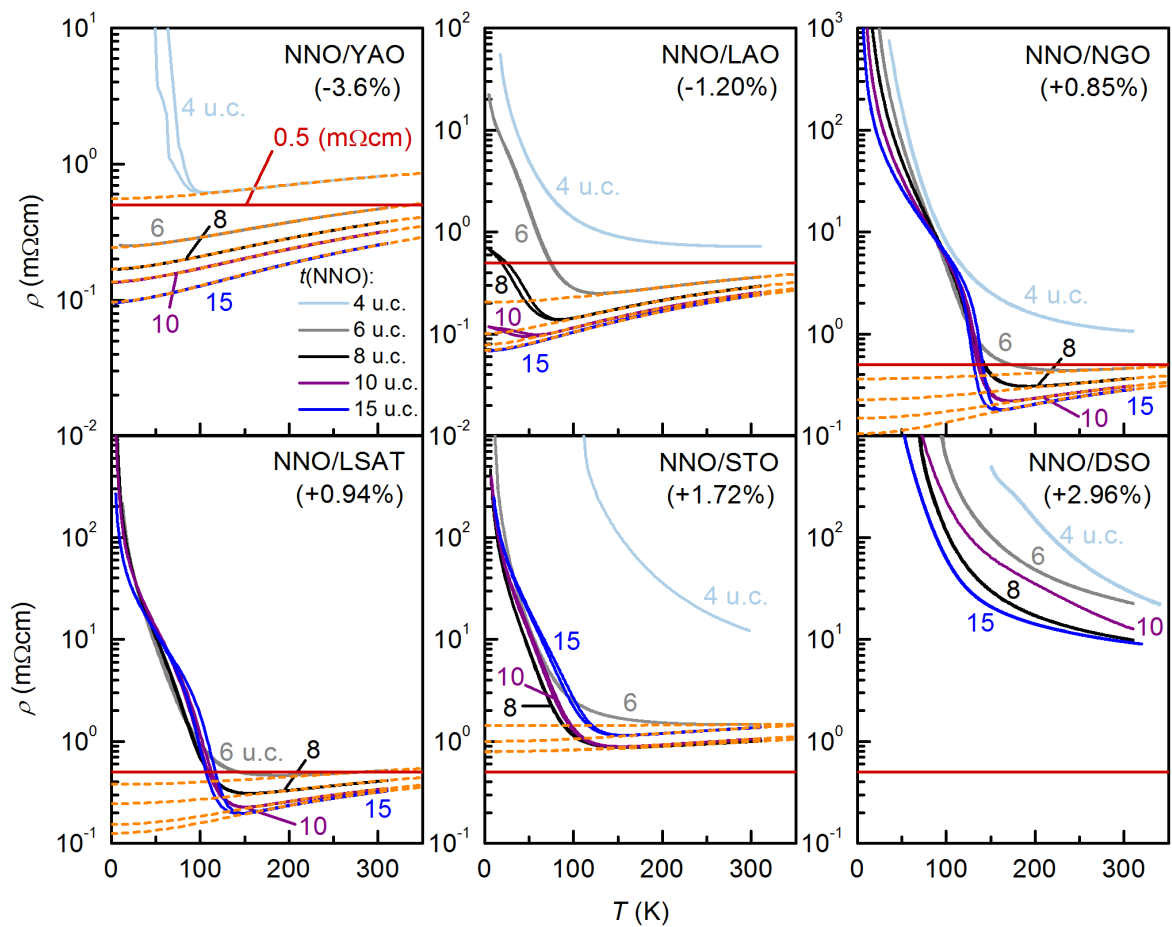

**Figure 1 | Metal-insulator transitions and temperature-dependence of the resistivity.** Shown is the resistivity as a function of temperature for NdNiO$_3$ films with thicknesses ranging between 4 and 15 u.c.'s on the six different substrates (YAO = YAlO$_3$, LAO = LaAlO$_3$, NGO = NdGaO$_3$, LSAT, STO = SrTiO$_3$, and DSO = DyScO$_3$). Each panel corresponds to a different substrate, with the corresponding epitaxial strain noted in parentheses. The solid lines are the experimental data and the dashed lines are fits using Eqs. (1) and (2). The horizontal solid line is the Mott-Ioffe-Regel limit according to Eq. (3).

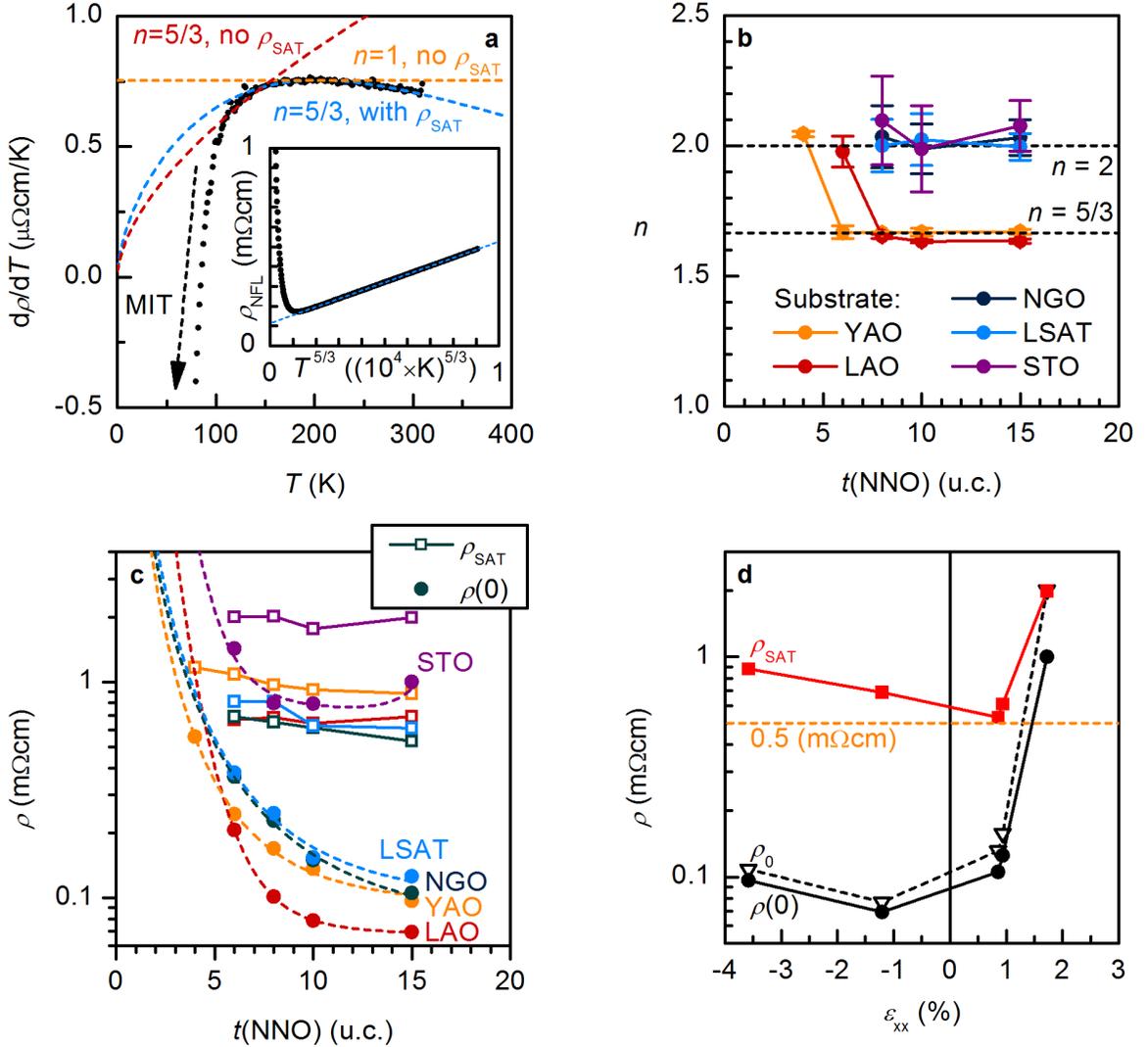

**Figure 2 | Resistance saturation and non-Fermi liquid behavior in the metallic state.** (a) Temperature derivative of the resistivity as a function of temperature (8 u.c. film on LaAlO$_3$). The negative slope at higher temperatures is a manifestation of $\rho_{SAT}$ in Eq. (2), and the dashed blue curve is a fit to Eq. (2) and $n = 5/3$. The downturn at lower temperatures is the MIT. The red and orange dashed curves are fits to Eqs. (1), with $n = 5/3$ and 1, respectively, which cannot describe the data. The inset shows a plot of $\rho_{NFL}$ (extracted from the fit to Eq. (2)) as a function of $T^{5/3}$. (b) Extracted exponents $n$ for different film thickness ($t_{(NNO)}$) and substrates. (c) $\rho_{SAT}$ and $\rho(0)$ as a function of $t_{(NNO)}$ for the different substrates. The dashed lines are extrapolated

polynomial fits to $\rho(0)$ used to determine the thicknesses correspond to $\rho_{SAT} = \rho(0)$ for the different substrates. (d) $\rho_{SAT}$, $\rho_0$ and $\rho(0)$ as a function of the in-plane epitaxial strain ($\varepsilon_{xx}$) for the 15 u.c. films. The dashed line is the Mott-Ioffe-Regel limit according to Eq. (3).

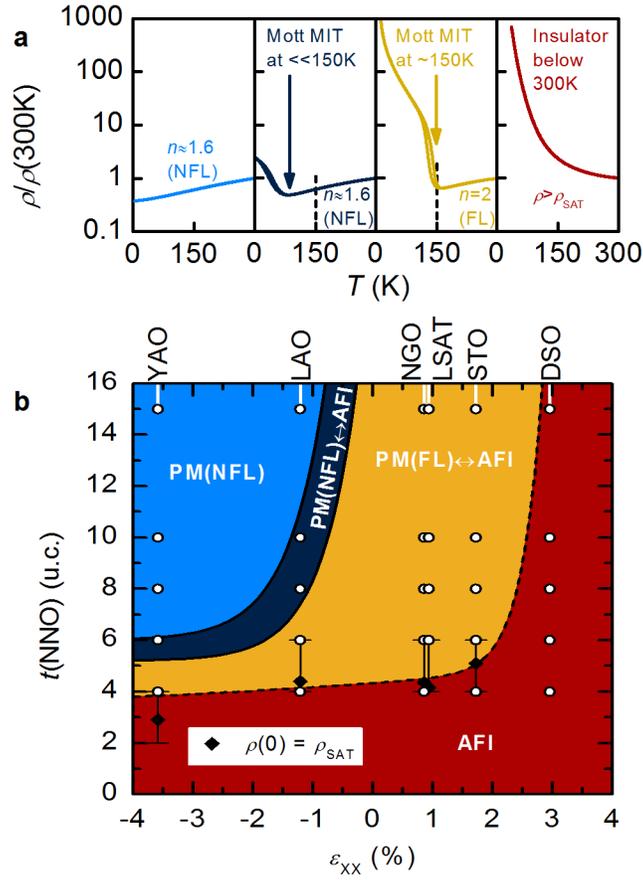

**Figure 3 | Strain-thickness phase diagram.** (a) Prototypes for the four basic behaviors seen in the $\rho$–$T$ curves shown in Fig. 1. (b) $\varepsilon_{xx}$ vs. $t$(NNO) phase diagram. The boundaries are drawn between four basic behaviors shown in (a). Each point indicates a transport curve in Fig. 1. The black diamonds are predictions for a metal-insulator transition based on $\rho(0) = \rho_{SAT}$.

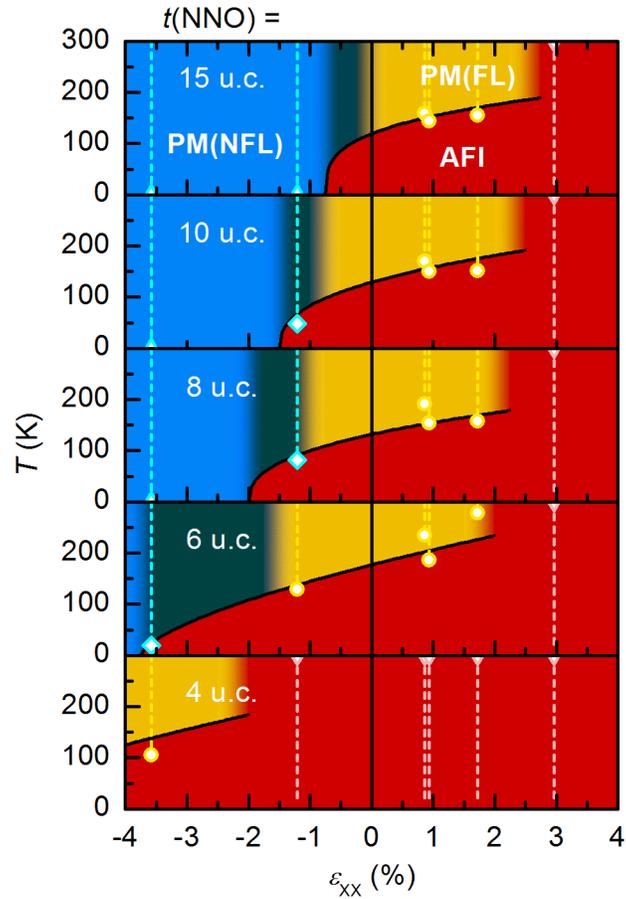

**Figure 4 | Strain-temperature phase diagram.** Each panel corresponds to a different NNO thickness. The symbols indicate the MIT temperatures measured for films under different strains. The colors of the regions correspond to those in Fig. 3, and boundaries are drawn to be consistent with Fig. 3.

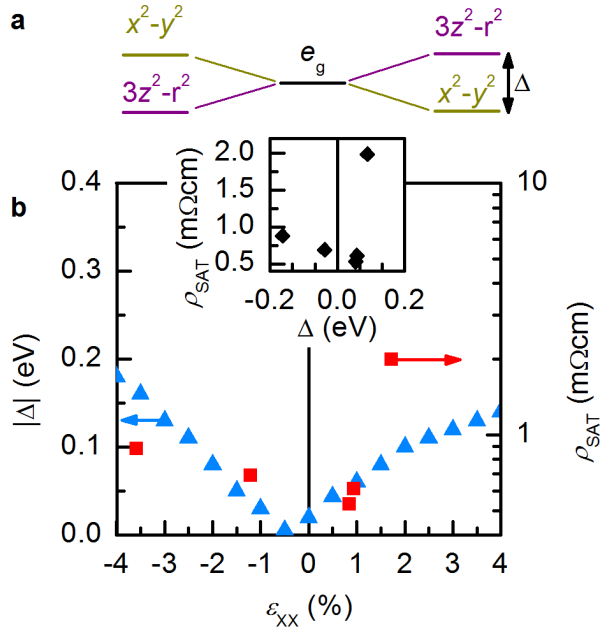

**Figure 5 | Saturation resistance and orbital splitting as a function of strain.** (a) Schematic showing the lifting of the $e_g$ orbital degeneracy in NdNiO$_3$. (b) Magnitude of the calculated orbital splitting $|\Delta|$ in NdNiO$_3$ and measured $\rho_{SAT}$ as a function of $\varepsilon_{xx}$. The inset illustrates the correlation between the two quantities.

# Tuning bad metal and non-Fermi liquid behavior in a Mott material: rare earth nickelate thin films

Evgeny Mikheev, Adam J. Hauser, Burak Himmetoglu, Nelson E. Moreno, Anderson Janotti, Chris G. Van de Walle, and Susanne Stemmer

Materials Department, University of California, Santa Barbara, CA 93106-5050, U.S.A.

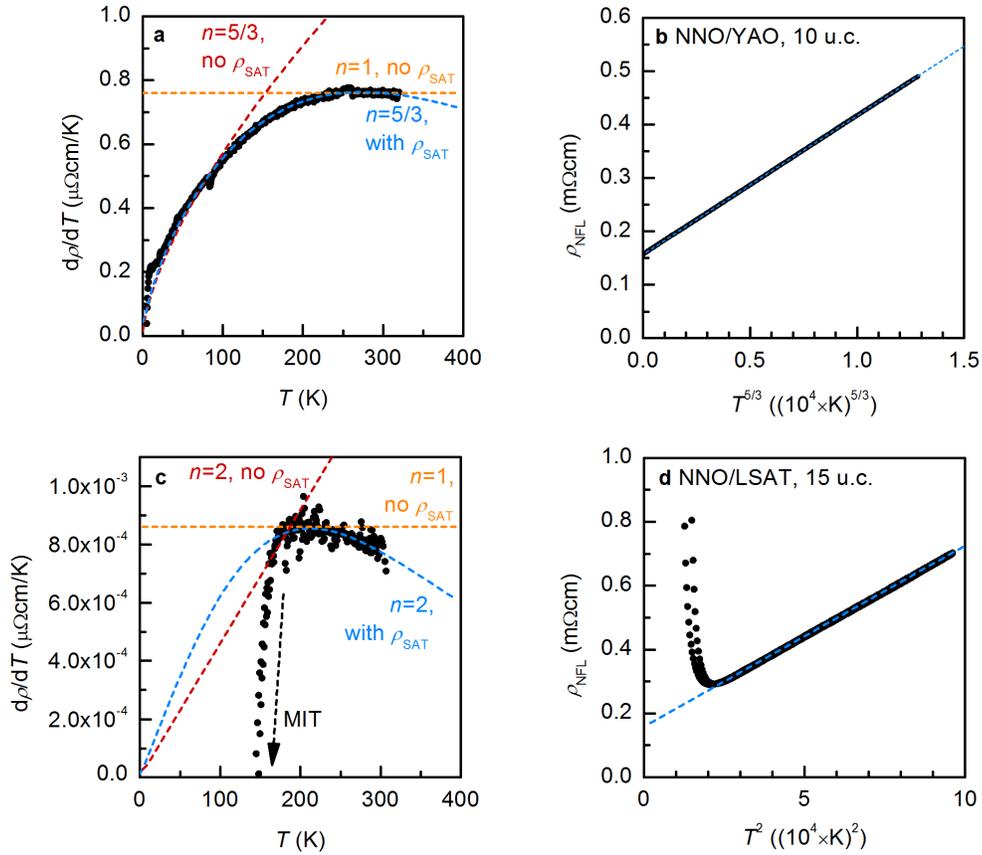

**Figure S1 | Resistivity as a function of temperature.** (a, c) Temperature derivative of the resistivity as a function of temperature for a 10 u.c. film on $YAlO_3$, a fully metallic non-Fermi liquid and a 15 u.c. thick film on LSAT, a LFL with a sharp metal-insulator transition near 150 K. The corresponding $\rho_{NFL}$ as a function of $T^n$ are shown in (b) and (d).

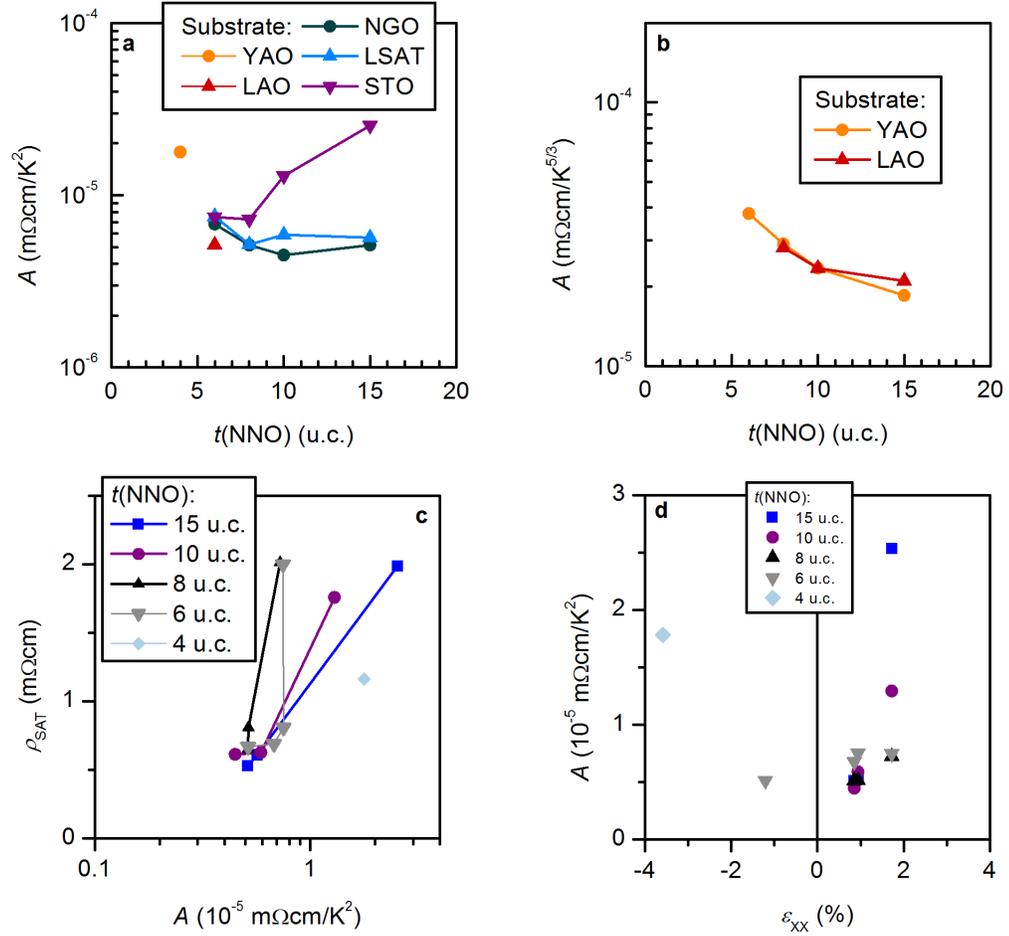

**Figure S2 | Electron-electron scattering coefficient *A*.** (a) *A* as a function of thickness for the LFLs (*n* = 2). (b) *A* as a function of thickness for all NFL films (*n* = 5/3). (c) Scaling between *A* and $\rho_{SAT}$, each curve corresponds to a specific NdNiO$_3$ thickness, $t_{NNO}$. (d) *A* as a function of strain for different film thicknesses.

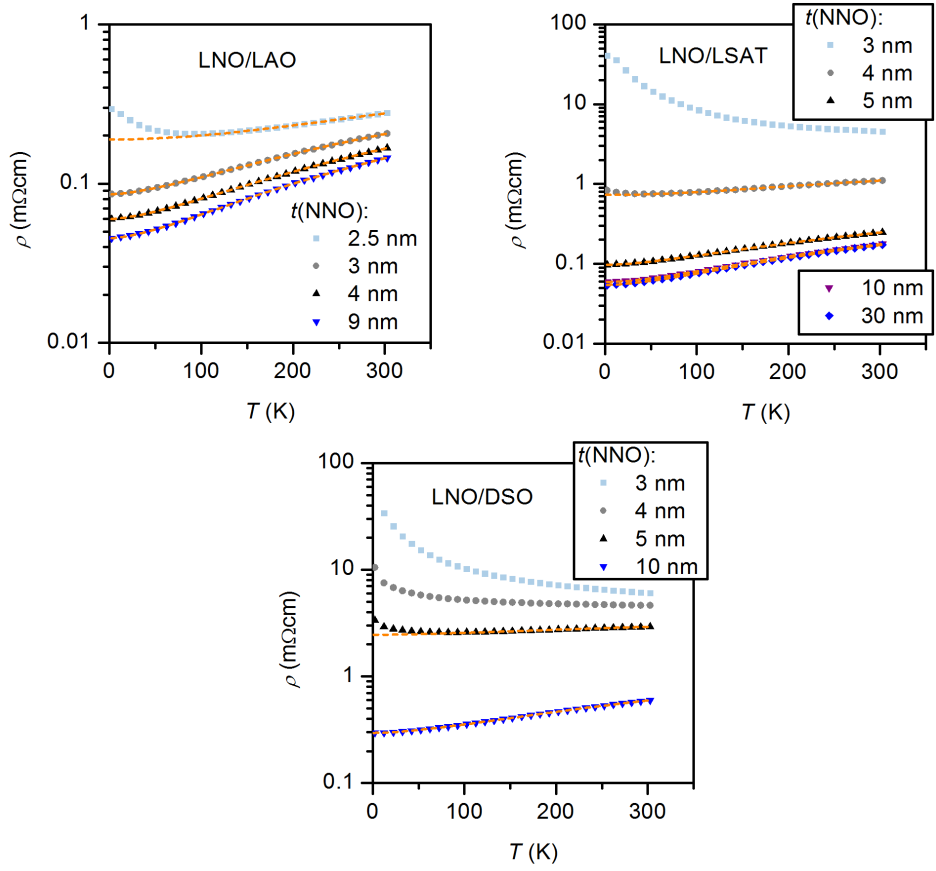

**Figure S3 | $\rho$-$T$ data for LaNiO$_3$.** Data re-plotted from ref. *(27)*. The dashed orange lines are fits to Eqs. (1) and (2).

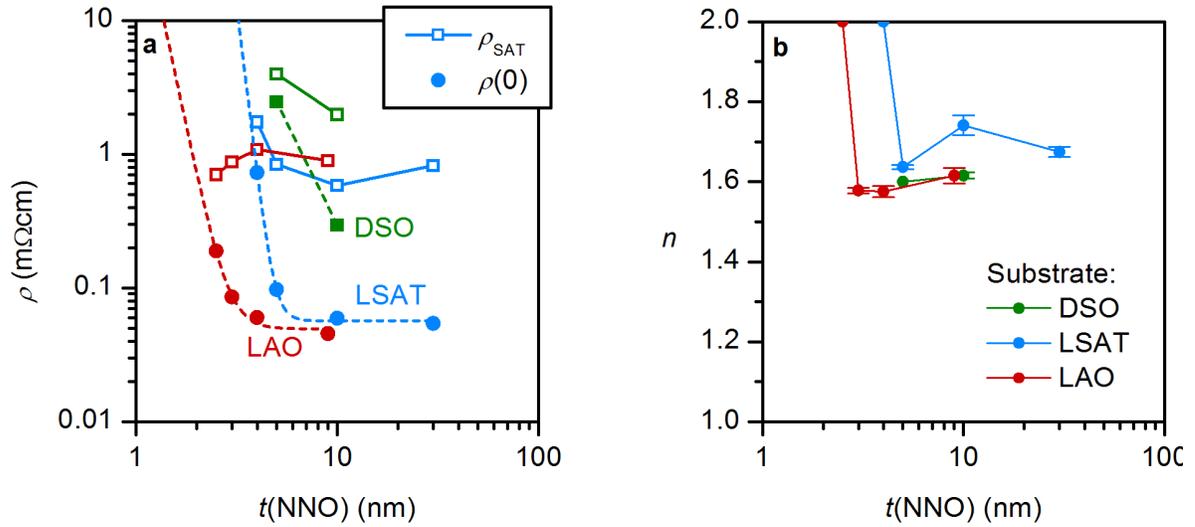

**Figure S4 | Saturation resistivity and NFL behavior in LaNiO$_3$.** (a) $\rho_{SAT}$ and $\rho(0)$ extracted from data in Fig. S3. The condition $\rho(0) = \rho_{SAT}$ accurately predicts the transition to an insulator at all temperatures. (b) Extracted exponent $n$, showing that thick LaNiO$_3$ films are non-Fermi liquids. All data are from ref. *(27)*.